\documentclass[10pt,a4paper]{article}
\usepackage{amsmath,amssymb}

\usepackage{changepage}

\usepackage[utf8x]{inputenc}

\usepackage{textcomp,marvosym}

\usepackage{nameref,hyperref}
\usepackage[sectionbib,square,numbers]{natbib}

\usepackage{microtype}
\DisableLigatures[f]{encoding = *, family = * }

\usepackage[table]{xcolor}

\usepackage{array}

\newcolumntype{+}{!{\vrule width 2pt}}

\newlength\savedwidth

\newcommand\thickhline{\noalign{\global\savedwidth\arrayrulewidth\global\arrayrulewidth 2pt}%
\hline
\noalign{\global\arrayrulewidth\savedwidth}}

\raggedright
\usepackage[aboveskip=1pt,labelfont=bf,labelsep=period,justification=raggedright,singlelinecheck=off]{caption}

\makeatletter
\renewcommand{\@biblabel}[1]{\quad#1.}
\makeatother

\date{}

\usepackage{lastpage,fancyhdr,graphicx}
\usepackage{epstopdf}
\DeclareMathOperator*{\argmin}{arg\,min}
\newcommand{\maxre}{\max \mathbf{r}_E}
\begin{document}
\vspace*{0.2in}

\begin{flushleft}
{\Large
\textbf\newline{A toolbox for rendering virtual acoustic environments in the
 context of audiology} % Please use "sentence case" for title and headings (capitalize only the first word in a title (or heading), the first word in a subtitle (or subheading), and any proper nouns).
}
\newline
\\
Giso Grimm\textsuperscript{1,2,*},
Joanna Luberadzka\textsuperscript{1},
Volker Hohmann\textsuperscript{1,2}
\\
\bigskip
\textbf{1} Medizinische Physik and Cluster of Excellence ``Hearing4all'',
Department of Medical Physics and Acoustics,
University of Oldenburg, Germany
\\
\textbf{2} H{\"o}rTech gGmbH, Marie-Curie-Str. 2, 26129 Oldenburg, Germany
\\
\bigskip

% Insert additional author notes using the symbols described below. Insert symbol callouts after author names as necessary.
% 
% Remove or comment out the author notes below if they aren't used.
%
% Primary Equal Contribution Note
%\Yinyang These authors contributed equally to this work.

% Additional Equal Contribution Note
% Also use this double-dagger symbol for special authorship notes, such as senior authorship.
%\ddag These authors also contributed equally to this work.

%\textcurrency Current Address: Dept/Program/Center, Institution Name, City, State, Country % change symbol to "\textcurrency a" if more than one current address note
% \textcurrency b Insert second current address 
% \textcurrency c Insert third current address


% Group/Consortium Author Note
%\textpilcrow Membership list can be found in the Acknowledgments section.

% Use the asterisk to denote corresponding authorship and provide email address in note below.
* g.grimm@uni-oldenburg.de

\end{flushleft}
\section*{Abstract}
A toolbox for creation and rendering of dynamic virtual acoustic
environments (TASCAR) that allows direct user interaction was
developed for application in hearing aid research and audiology.
This technical paper describes the general software structure and the
time-domain simulation methods, i.e., transmission model, image source
model, and render formats, used to produce virtual acoustic
environments with moving objects.
Implementation-specific properties are described, and the
computational performance of the system was measured as a function of
simulation complexity.
Results show that on commercially available commonly used hardware the
simulation of several hundred virtual sound sources is possible in the
time domain.\footnote{Parts of this study have been presented at the
  Linux Audio Conference, Mainz, Germany, 2015.}

% Please keep the Author Summary between 150 and 200 words
% Use first person. PLOS ONE authors please skip this step. 
% Author Summary not valid for PLOS ONE submissions.   
%\section*{Author summary}
%Lorem ipsum dolor sit amet, consectetur adipiscing elit. Curabitur eget porta erat. Morbi consectetur est vel gravida pretium. Suspendisse ut dui eu ante cursus gravida non sed sem. Nullam sapien tellus, commodo id velit id, eleifend volutpat quam. Phasellus mauris velit, dapibus finibus elementum vel, pulvinar non tellus. Nunc pellentesque pretium diam, quis maximus dolor faucibus id. Nunc convallis sodales ante, ut ullamcorper est egestas vitae. Nam sit amet enim ultrices, ultrices elit pulvinar, volutpat risus.

%\linenumbers

% Use "Eq" instead of "Equation" for equation citations.
\section*{Introduction}
Hearing aids are evolving from simple amplifiers to complex signal
processing devices. Current hearing devices typically contain
spatially sensitive algorithms, e.g., directional microphones,
direction of arrival estimators, or binaural noise reduction, as well
as automatic classification of the acoustic environment that is used
for context-adaptive processing and amplification
\cite{Hamacher2005}.
Several of these features cannot be tested in the current lab-based
setups for hearing-aid evaluation, because they employ rather simple
acoustic configurations.
Furthermore, it was shown in several studies that hearing aid
performance depends on the spatial complexity of the environment, and
that the hearing aid performance in simple laboratory conditions is
not a good predictor of the performance in more realistic environment
or in the real life
\cite{Ricketts2000,Cord2004,Bentler2005,Best2015,Grimm2016}.
Finally, recent developments in hearing aid technology led to an
increased level of interaction between the user, the environment and
the hearing devices, e.g., by means of motion interaction
\cite{Tessendorf2011a,Tessendorf2011b}, gaze direction
\cite{Kidd2013} or even with brain-computer interfaces
\cite{deVos2014}.
Thus, for an improved assessment of hearing aid benefit as well as for the
development and evaluation of user interaction techniques, a
reproduction of complex listening environments in the laboratory may
be beneficial.

Advances in computer technology in combination with recent
multi-channel reproduction \cite{Berkhout1993,Pulkki1997,Daniel2001}
and acoustic simulation methods \cite{Allen1979,Lentz2007,Wendt2014b}
allow for the reproduction of virtual acoustic environments in the
laboratory.
Limitations in reproduction and simulation quality have been studied
in terms of perceptual effects \cite{Bertet2013,Grimm2015c} as well
as in terms of technical accuracy of hearing aid benefit prediction
\cite{Grimm2015b,Oreinos2015}.
These studies support the general applicability of virtual acoustic
environments to hearing-aid evaluation and audiology, but show that
care must be taken in designing the simulation and reproduction
methods, to ensure that the outcome measures are not biased by the
artifacts of the applied methods.

Several further requirements apply when using virtual acoustic
environment in hearing research and audiology.
To allow for a systematic evaluation of hearing device performance,
virtual acoustic environments need to be reproducible and scalable in
their complexity.
The presence of early reflections and late reverberation in the
simulation is essential for the application of hearing aid evaluation,
since both of these factors may affect hearing aid performance
\cite{Ricketts2000}.
For assessment of user interaction, but also for the analysis of
hearing aid benefit, simulation of the effects of motion of listeners
and sources might be desired. These effects do not only include
time-variant spatial cues, but also Doppler-shift and time-variant
spectral cues due to comb filtering.

Existing virtual acoustic environment engines often target authentic
simulations for room acoustics \cite{EASE,Naylor1993},
resulting in a large computational complexity.
They typically render impulse responses for off-line analysis or
auralization and thus do not allow studying motion and user
interaction.
Other interactive tools, e.g., the SoundScapeRenderer
\cite{Ahrens2008}, do not provide all features required here, such as
room simulation and diffuse source handling.

To accommodate the requirements listed above, a toolbox for acoustic
scene creation and rendering (TASCAR) was developed as a Linux audio
application \cite{Grimm2015a} with an open source core
\cite{tascargit} and commercial support \cite{tascarhtch}.
The aim of TASCAR is to interactively render complex and time varying
virtual acoustic environments via loudspeakers or headphones.
For a seamless integration into existing measurement tools of
psycho-acoustics and audiology, low-delay real-time processing of
external audio streams in the time domain is applied, and interactive
modification of the geometry is possible.
This technical paper aims at describing the general structure of
applications in hearing aid evaluation and audiology, the applied
underlying simulation and rendering methods, and their specific
implementation.
A measurement of the computational performance and its underlying
factors is provided to allow for an estimation of maximum simulation
complexity in relation to the available computing power.
This paper also serves as a technical reference for the TASCAR open
source software (TASCAR/GPL).

\section*{General structure} 

\begin{figure}[!h]
%\centerline{\includegraphics[width=150mm]{figures/components_main}}
\centerline{\includegraphics[width=150mm]{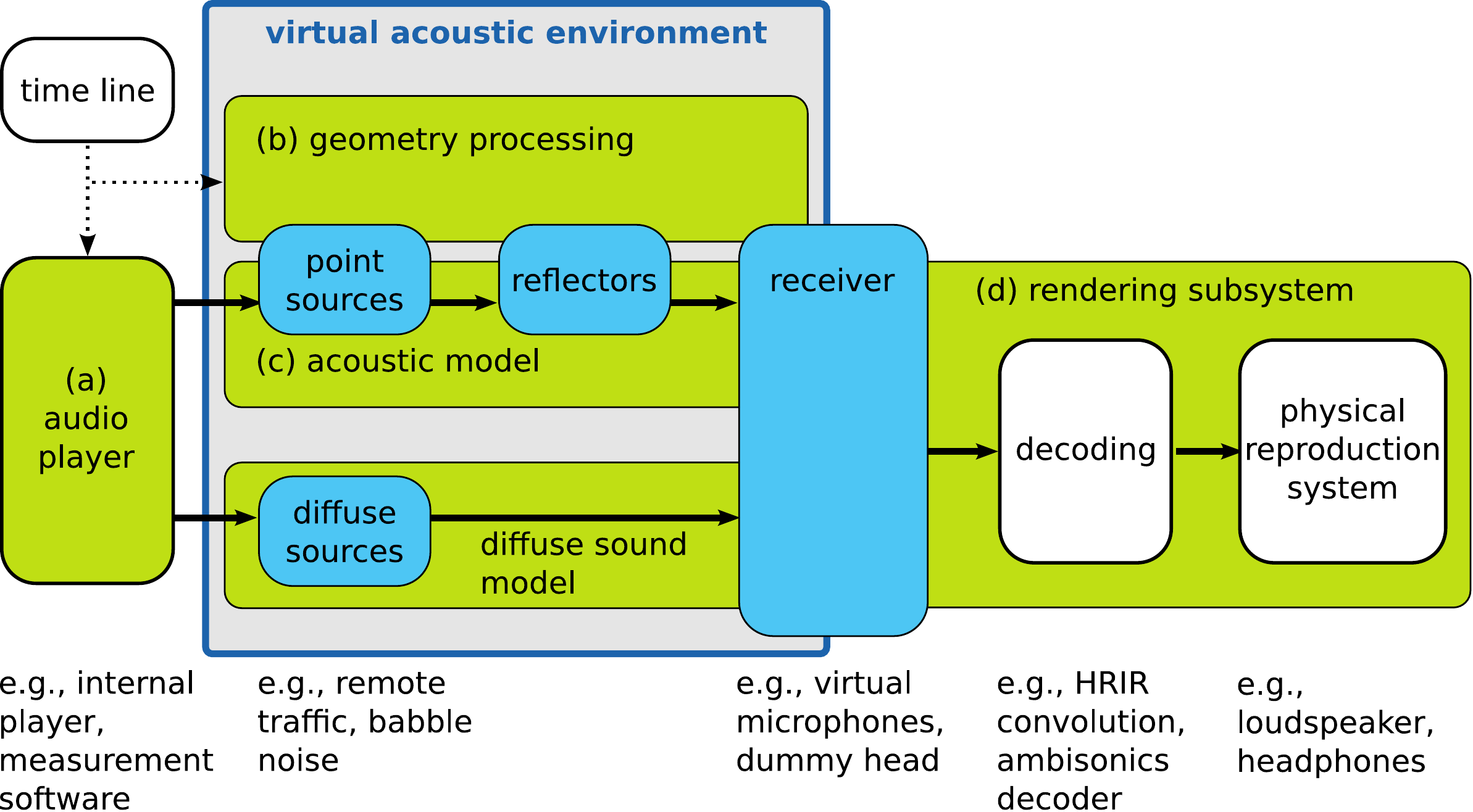}}
\caption{{\bf The major components of TASCAR are the audio player (a), the
  geometry processor (b), the acoustic model (c) and the rendering
  subsystem (d). Point sources and diffuse sources are the interface
  between the audio player and the acoustic model. Receivers are the
  interface between the acoustic model and the rendering
  subsystem.}}\label{fig:components}
\end{figure}

The structure of TASCAR can be divided into four major components
(see Figure \ref{fig:components} for an overview):
The audio player (block a in Fig \ref{fig:components}) serves as a
source of audio signals.
The geometry processor (block b) controls position and orientation of
objects over time.
The acoustic model (blocks c) simulates sound propagation, room
acoustics and diffuse sounds.
Finally, the rendering subsystem (block d) renders the output of the
acoustic model for a physical reproduction system.

A virtual acoustic environment in TASCAR is defined as a space
containing several types of objects: point sources (e.g., speakers,
distinct noise sources), diffuse sources (e.g., remote traffic, babble
noise), receivers (e.g., dummy head), reflectors (e.g., boundaries
of a room) and obstacles.
Source objects are provided with the audio content, delivered either
by the internal audio player module, or externally e.g., from physical
sources, audiological measurement tools, or digital audio workstations
(DAW).

Objects in the virtual acoustic scene can change their positions and
orientations over time.
Information about the object geometry at a given time is taken either
from sampled trajectories, from algorithmic trajectory generators, or
from external devices, e.g., a joystick or head-motion tracker
(interactive controller of an object's movement, e.g., motion of a
dummy head).

Geometry information is exploited in the acoustic model to modify the
input audio signals delivered by the audio player.
Modifications performed by the acoustic model mimic basic acoustic
properties like distance law, reflections and air absorption.
The resulting sound corresponds to the time-variant spatial
arrangement of the objects in the virtual scene.
Geometry data can also be exchanged with external modules, e.g., game
engines, to make the visualization consistent with the acoustic scene
content.

At the final stage of the acoustic model, there is a receiver model,
which encodes the modified signals into a receiver type specific
render format, used subsequently by the rendering subsystem for the
reproduction of the simulated environment on a physical reproduction
system.

\section*{Simulation methods}

\subsection*{Geometry processing}\label{sec:geometry}

Each object in a virtual acoustic environment is determined by its
position $\mathbf{p}(t)$ and orientation $\mathbf{\Omega}(t)$ in space
at a given time $t$.
Position is defined in Cartesian coordinates
$\mathbf{p}=(p_x,p_y,p_z)$, and orientation is defined in the Euler
angles, $\mathbf{\Omega}=(\Omega_z,\Omega_y,\Omega_x)$, where
$\Omega_z$ is the rotation around the z-axis, $\Omega_y$ around the
y-axis and $\Omega_x$ around the x-axis.

Trajectories $\mathbf{\Gamma}$ for a moving object are created by specifying
the position and orientation for more then one point in time:
\begin{eqnarray*}
\mathbf{\Gamma}_{\mathbf{p}} &=& \left\{  \mathbf{p}(t_1), \mathbf{p}(t_2), \mathbf{p}(t_3), \dots \right\}\\
\mathbf{\Gamma}_{\mathbf{\Omega}} &=& \left\{ \mathbf{\Omega}(t_1), \mathbf{\Omega}(t_2), \mathbf{\Omega}(t_3), \dots \right\},
\end{eqnarray*}
where ${t}_{1},{t}_{2},\dots\in\mathbb{R}$ are the sampling times of
the trajectory.
The time variant position $\mathbf{p}(t)$ is linearly interpolated
between sample times of $\mathbf{\Gamma}_{\mathbf{p}}$, either in Cartesian
coordinates, or in spherical coordinates relative to the origin,
respectively.
The time variant orientation $\mathbf{\Omega}(t)$ is linearly
interpolated from $\mathbf{\Gamma}_\Omega$, in Euler-coordinates.
To apply the orientation to objects, a rotation matrix $\mathbf{O}$ is
calculated from the Euler coordinates.

\subsection*{Acoustic model}

For each sound source object $k$, the acoustic model modifies its
associated original source signal $x(t)$ delivered by the audio player
using geometry data into an output signal $y(t)$ that is then used as
input signal to a receiver.
The performed computations simulate basic acoustic phenomena as
described below.
Signals $y(t)$ serve at the subsequent stage to calculate the output
of a receiver (see Section on render formats below).

The acoustic model consists of the source model (omni-directional or
frequency-dependent directivity), the transmission model simulating
sound propagation, an image source model, which depends on the
reflection properties of the reflecting surfaces as well as on the
`visibility' of the reflected image source, and a receiver model,
which encodes the direction of the sound source relative to the
receiver into an receiver output for further processing by the
rendering subsystem.

\subsubsection*{Image source model}

\begin{figure*}[!h]
\includegraphics[width=\columnwidth]{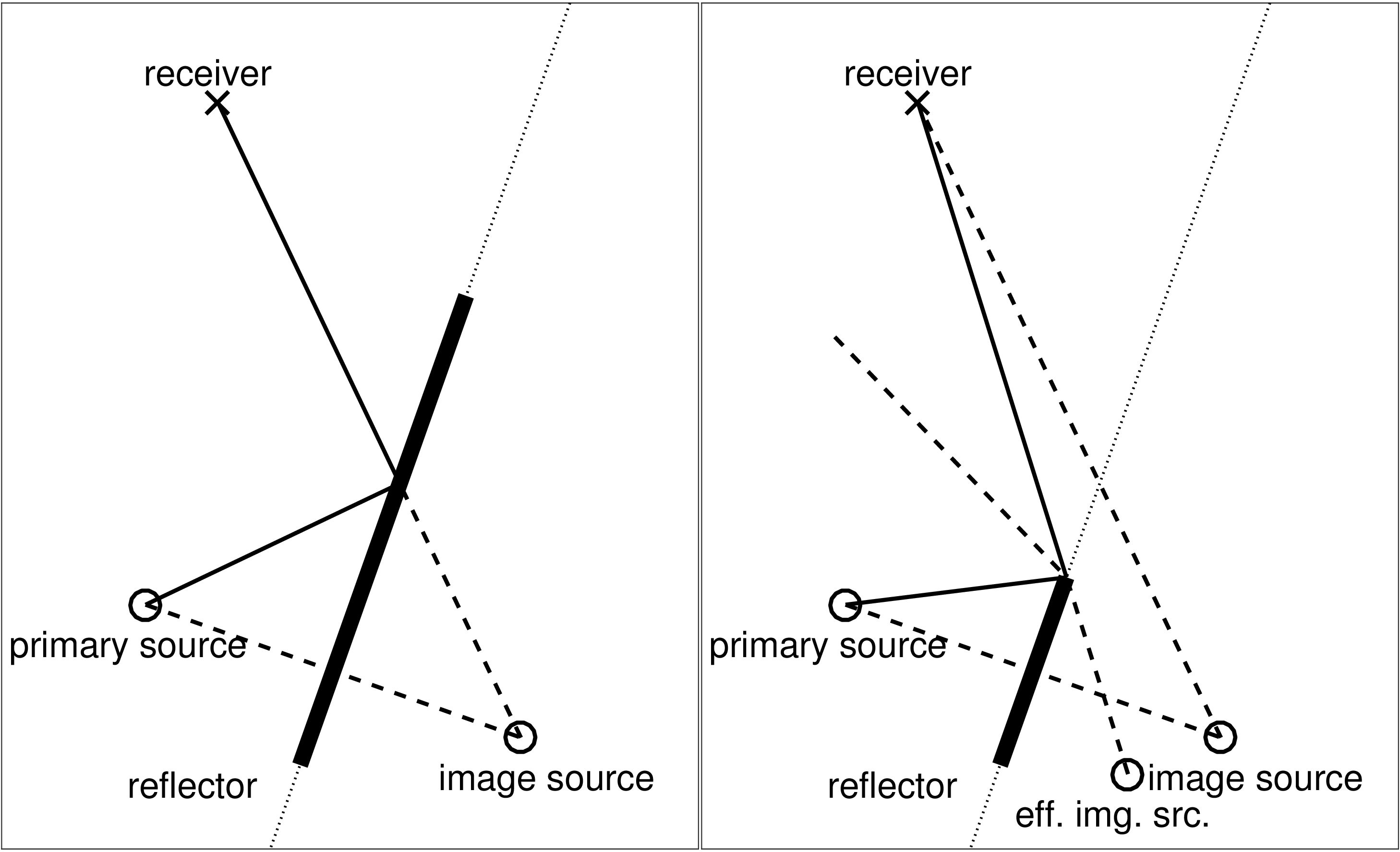}%
\caption{{\bf Schematic sketch of the image model geometry. Left panel:
  `specular' reflection, i.e., the image source is visible within
  the reflector; right panel: `edge' reflection.}}\label{fig:imagesrc}
\end{figure*}

Early reflections are generated with a geometric image source model,
i.e., reflections are simulated for each reflecting plane surface with
polygon-shaped boundary by placing an image source at the position of
the reflection.
Each image source is rendered in the time domain, in the same way as
primary sources.
This is different to the more efficient ``shoe-box'' image source
models commonly used in room acoustic simulations \cite{Allen1979},
which calculate impulse responses by solving the wave equations.
For a first order image source model, each pair of primary source and
reflector face creates an image source, where the plane on which the
reflector lies is a symmetry axis between the primary and image source
(see Fig \ref{fig:imagesrc}).
The image source position $\mathbf{p}_{img}$ is determined by the
closest point on the (infinite) reflector plane $\mathbf{p}_{cut}$ to
the source $\mathbf{p}_{src}$: $\mathbf{p}_{img} =
2\mathbf{p}_{cut}-\mathbf{p}_{src}$.

For higher order image source models, lower order image sources are
treated as primary sources leading to
higher order image sources.

%REFLECTION TYPES:
The image source position itself is independent of the receiver
position. However, for finite reflectors there are two types of
reflections in TASCAR, and depending on the receiver position it is
determined which reflection type is executed (see Fig
\ref{fig:imagesrc}).
%
% visible = specular:
If the intersection point $\mathbf{p}_{is}$ of the line connecting the
image source with the receiver and the reflector plane lies within the
reflector boundaries, the image source is `visible' in the
reflector, and a `specular' reflection is applied.
% invisible = edge:
If $\mathbf{p}_{is}$ is not within the reflector boundaries, the
source is `invisible' from a receiver perspective and the `edge
reflection' is applied.
% apparent image source position for edge:
For `edge' reflections, the apparent image source position is shifted
so that the distance between the source and receiver remains
unchanged, whereas the receiver, edge of the reflector and the apparent
source position form one line (see Fig \ref{fig:imagesrc}, right panel).
The angle $\theta$ by which the image source is shifted to create
effective image source controls a soft-fade gain by which the source signal is
multiplied $g$: 
\begin{equation}
g = \cos(\theta)^{\kappa}
\end{equation}
The coefficient $\kappa=2.7$ was chosen for a rough approximation of
diffraction of speech-shaped signals and medium-sized reflectors.
If a receiver or a sound source are behind the reflector, the image source
is not rendered. A reflector object has only one reflecting side in
the direction of the face normal.  

%Reflection properties:
To simulate the reflection properties of a reflector object, the
source signal is filtered with a first order low pass filter
determined by a reflectivity coefficient $\rho$, and a damping
coefficient $\delta$, which can be specified for each reflector
object:
\begin{equation}
y(t) = \delta y(t-f_s^{-1}) + \rho x(t).
\end{equation}
In room acoustics material properties are commonly defined by
frequency dependent absorption coefficients $\alpha(f)$.
These can be calculated from the reflection filter coefficients $\rho$
and $\delta$ by
\begin{equation}
\alpha(f)=\left(1-\left|\rho\frac{1-\delta}{1-\delta e^{-i 2\pi f f_s^{-1}}}\right|\right)^2.
\end{equation}
The filter coefficients $\rho$ and $\delta$ can be derived from
frequency dependent absorption coefficients by minimization of the
mean-square error between desired $\tilde\alpha(f)$ and $\alpha(f)$ derived
from the filter coefficients.

\subsubsection*{Source directivity}

For the simulation of source directivity, the receiver position
relative to the source
\begin{equation}
\mathbf{p}_{rec,rel} = \mathbf{O}_{src}^{-1} (\mathbf{p}_{rec}-\mathbf{p}_{src})
\end{equation}
is calculated.
% Omni-directional sources ignore the value of $\mathbf{p}_{rec,rel}$.
%
Frequency-dependent directivity with omni-directional characteristics
at low frequencies and higher directivity at high frequencies is
achieved by controlling a low-pass filter by the angular distance
between the receiver and the source direction.
The normalized relative receiver position $\tilde{\mathbf{p}}_{rec,rel}$ is
\begin{equation}
\tilde{\mathbf{p}}_{rec,rel} = \frac{\mathbf{p}_{rec,rel}}{||\mathbf{p}_{rec,rel}||}
\end{equation}
The cosine of the angular distance is then $\tilde{p}_{x,rec,rel}$.
The cut-off frequency $f_{6dB}$ defines the frequency, for which
$-6$\,dB at $\pm 90$\,degrees are achieved.
With $\xi = \frac{\pi f_{6dB}/f_s}{\log( 2 )}$, a first order low-pass filter with the recursive filter coefficient $c_{lp}$,
\begin{equation}
  c_{lp} = \left(\frac12 - \frac12 \tilde{p}_{x,rec,rel}\right)^{\xi (f_{cut})},
\end{equation}
is applied to the signal, to achieve the frequency-dependent
directivity, or in other words, the direction-dependent frequency
characteristics.

\subsubsection*{Transmission model}

The transmission model simulates the delay, attenuation and air
absorption, which depend on the distance $r(t)$ between the sound
source (primary or image source) and the receiver, as well as attenuation,
caused by obstacles between source and receiver.
Point sources follow a $1/r$ sound pressure law, i.e., doubling the
distance $r$ results in half of the sound pressure. 
% Air absorption:
Air absorption is approximated by a simple first order low-pass filter
model with the filter coefficient $a_1$ controlled by the distance:
\begin{equation}
a_1 = e^{-\frac{r(t) f_{s}}{c\alpha}},
\end{equation}
where $f_s$ is the sampling frequency and $c$ the speed of sound.
The empiric constant $\alpha=7782$ was manually adjusted to provide
appropriate values for distances below 50 meters.
This approach is very similar to that of \cite{Huopaniemi1997} who
used an FIR filter to model the frequency response at certain
distances. However, in this approach the distance parameter $r$ can be
varied dynamically.
The distance dependent part of the transmission model without
obstacles can then be written as
\begin{equation}
y(t) = a_1 y(t-f_s^{-1}) + (1-a_1)%\underbrace{}_\textrm{air absorption} 
\frac{ x(t-r(t)c^{-1})}{r(t)}%\underbrace{}_\textrm{attenuation and delay}
,
\end{equation}
where $x(t)$ is the source signal at time $t$, and $y(t)$ is the
output audio signal of the transmission model.
The time-variant delay line uses either nearest neighbor interpolation
or sinc interpolation, depending on the user needs and computational
performance of the computing system.

Obstacles are modeled by plane surfaces with polygon-shaped
boundaries.
The acoustic signal is split into a direct path, which is attenuated
by the obstacle-specific frequency-independent attenuation $a_o$, and
an indirect path, to which a simple diffraction model is applied.
The diffracted path is filtered with a second order low pass filter
which is controlled by the shortest path from the source via the
obstacle boundary to the receiver. 
With the angle $\theta_o$ between the connection from the intersection
point of the shortest path with the obstacle boundary to the source
position, and the connection from the receiver position to the
intersection point, the cut-off frequency of the low-pass filter is
\begin{equation}
f_o = 3.8317\frac{c}{2\pi a \sin(\theta_o)},
\end{equation}
with the aperture $a=2\sqrt{A/\pi}$ defined as the radius of a circle
with the same area $A$ as the obstacle polygon.
This simple diffraction model is based on the diffraction on the
boundary of a circular disc \cite{Airy1835}, however,
position-dependent notches are not simulated.
The diffracted signal is weighted with $1-a_o$ and added to the
attenuated signal.

\subsubsection*{Diffuse sources}

Sound sources with lower spatial resolution, e.g., diffuse background
noise or diffuse reverberation \cite{Wendt2014b}, are added
in first order Ambisonics (FOA) format.
No distance law is applied to these sound sources; instead, they have
a rectangular spatial range box, i.e., they are only rendered if the
receiver is within their range box, with a von-Hann ramp at the
boundaries of the range box.
Position and orientation of the range box can vary with time.
The size of the range box is typically adjusted to match the dimension
of the simulated room.
The diffuse source signal is rotated by the difference between
box orientation and receiver orientation.

Diffuse reverberation is not simulated in TASCAR.
To use diffuse reverberation, the input signals of the image source
model can be passed to external tools which return FOA signals, e.g.,
feedback-delay networks or convolution with room impulse responses in
FOA format \cite{openairlib}.
A smooth transition between early reflections from the image source
model and diffuse reverberation based on room impulse responses can be
achieved by removing the first reflections from the impulse responses.
To account for position-independent late reverberation, room receivers
can render independent from the distance between source and receiver,
e.g., the transmission model can be replaced by a room-volume
dependent fixed gain.

\subsubsection*{Receiver model}

The interface between the acoustic model and the rendering subsystem
is the receiver.
A receiver renders the output of the transmission model depending on
the relative position and orientation between receiver and sound
source.
Signals from the transmission models belonging to all sound sources
are summed up after direction-dependent processing.
The render format determines the number of channels and the method of
encoding the relative spatial information into a multi-channel audio
signal.
The output signal of a receiver is
\begin{equation}
\mathbf{z}(t)=\left(z_1\left(t\right),z_2\left(t\right),\dots,z_N\left(t\right)\right).
\end{equation}
The receiver functionality can be split into the {\em panning} or
directional encoding of primary and image sources $y(t)$, and the {\em
  decoding} of diffuse source signals $\mathbf{f}(t)$ in first order
Ambisonics format with Furse-Malham normalization ('B-format'):
\begin{equation}
\mathbf{z}(t) = 
\underbrace{\sum_{k=1}^K \mathbf{w}(\mathbf{p}_{rel,k}) y_k(t)}_\textrm{panning} + 
\underbrace{\sum_{l=1}^L \mathbf{D}\hat{\mathbf{O}}_{rec}^{-1}\mathbf{f}_l(t)^T}_\textrm{diffuse decoding}
\label{eq:pandec}
\end{equation}
%panning part
In the panning part, the driving weights
$\mathbf{w}=(w_1,w_2,\dots,w_N)$ depend on the direction of the
relative source position in the receiver coordinate system,
$\mathbf{p}_{rel,k}=\mathbf{O}_{rec}^{-1}\left(\mathbf{p}_{k}-\mathbf{p}_{rec}\right)$;
$\mathbf{O}_{rec}$ is the receiver orientation matrix, and
$\mathbf{p}_{k}$ is the position of the $k$-th sound source.
$y_k(t)$ is the output signal of the transmission model, i.e., it
contains the distance-dependent gain, air absorption and obstacle
attenuation, for the $k$-th source; $K$ is the number of all primary
and image point sources.

%diffuse part
In the diffuse decoding part, $\mathbf{D}$ is the receiver-type
specific first order Ambisonics decoding matrix for the $w$, $x$, $y$
and $z$ channels of the first order Ambisonics signal,
\begin{equation*}
\mathbf{D} = \left(
\begin{array}{cccc}
  d_{1,w} & d_{1,x} & d_{1,y} & d_{1,z} \\
  \vdots    & \vdots  & \vdots  & \vdots  \\
  d_{n,w}  & d_{n,x} & d_{n,y} & d_{n,z}
\end{array}\right)\textrm{,}
\end{equation*}
and $\hat{\mathbf{O}}_{rec}^{-1}$is the rotation matrix for first
order Ambisonics signals, to compensate the receiver orientation.
$\mathbf{f}_l$ is the first order Ambisonics signal of the $l$-th
diffuse source, rotated by the source orientation; $L$ is the number
of all diffuse sources, including diffuse reverberation inputs.

\subsection*{Render formats}\label{sec:receivermodel}

The render formats of TASCAR  can be divided into three categories:
{\em Virtual microphones} simulate the characteristics of microphones.
They primarily serve as a sensor in a virtual acoustic environment.
{\em Speaker-based} receiver types render signals which can drive real
or virtual loudspeakers, used for auralization of virtual scenes.
{\em Ambisonics} receiver types render the scenes to first, second or
third order Ambisonics format, which can be rendered to virtual
microphones, loudspeakers or other reproduction methods using external
decoders.
Receivers can render either for three-dimensional reproduction or for two-dimensional reproduction. 
In both cases, the directional information of the relative source position is encoded in the normalized relative source position, 
\begin{equation}
\tilde{\mathbf{p}}_{rel} =
\frac{\mathbf{p}_{rel}}{||\mathbf{p}_{rel}||}.
\end{equation}
However, in the two-dimensional case \(\mathbf{p}_{rel}\) is projected
onto \(x,y\)-plane before the normalization by setting its
\(z\)-component to zero. In both cases, the acoustic model, containing
all distance-dependent effects, and the image source model are
calculated based on the three-dimensional relative source position.

\subsubsection*{Virtual microphones} 
The virtual microphone receiver type has a single output channel. The
driving weight is
\begin{equation}
w = 1 + a (\tilde{p}_{rel,x} - 1).
\end{equation}
It's directivity pattern can be controlled between omni-directional and
figure-of-eight with the directivity coefficient $a$; with $a=0$ this
is an omni-directional microphone, with $a=\frac12$ a standard
cardioid, and with $a=1$ a figure-of-eight.
The diffuse decoding matrix is
\begin{equation}
\mathbf{D} = \left( \sqrt{2}(1-a) \quad a \quad 0\quad  0 \right).
\end{equation}
The factor \(\sqrt{2}\) of the \(w\)-channel is needed to account for
the Furse-Malham normalization of the diffuse signals.

\subsubsection*{Speaker-based render formats}
This class of render formats contains all types which render the signals
directly to a loudspeaker array.
The number $N$ and position $\mathbf{s}_n$ of speakers can be
user-defined; $\tilde{\mathbf{s}}_n$ is the normalized speaker
position.
A measure of angular distance between a source and a loudspeaker is
$d_n = 1-\tilde{\mathbf{s}}_n \tilde{\mathbf{p}}_{rel}^T$.
%Nearest speaker
The most basic speaker-based receiver type is {\em nearest speaker
  panning} (NSP).
The driving weights are:
\begin{equation}
w_n = \left\{
\begin{array}{ll}
1 & n=\argmin\{d_n\} \\ 
0 & \textrm{otherwise}\\
\end{array}
\right.
\end{equation}
%VBAP
Another commonly used speaker-based render format is two-dimensional  {\em vector-base
amplitude panning} (VBAP) \cite{Pulkki1997}. The two speakers $n_1$ and $n_2$ which are
closest to the source are chosen.
A gain vector $(g_{n_1},g_{n_2})^T$ based on the normalized speaker
positions and the normalized relative source
position in the $x,y$-plane is defined:
\begin{equation}
\left(
\begin{array}{c}
g_{n_1}\\
g_{n_2}
\end{array}
\right)
 = \left(
\begin{array}{cc}
\tilde{s}_{n_1,x} & \tilde{s}_{n_2,x}\\
\tilde{s}_{n_1,y} & \tilde{s}_{n_2,y}\\
\end{array}
\right)^{-1}\left(
\begin{array}{c}
\tilde{p}_{rel,x}\\
\tilde{p}_{rel,y}\\
\end{array}
\right)
\end{equation}
Then the driving weights are
\begin{equation}
\left(
\begin{array}{c}
w_{n_1}\\
w_{n_2}
\end{array}
\right)
 = \frac{1}{\sqrt{g_{n_1}^2+g_{n_2}^2}}
\left(
\begin{array}{c}
g_{n_1}\\
g_{n_2}\\
\end{array}
\right).
\end{equation}
For ambisonic panning with arbitrary order, the signal of each source
is encoded into horizontal Ambisonics format (HOA2D).
Decoding into speaker signals is applied after a summation of the
signals across all sources.
In the decoder, the order gains can be configured to form a 'basic'
decoder or a '$\maxre$' decoder \cite{Daniel2001}.
An equal circular distribution of loudspeakers is assumed for this
render format.
Although this receiver applies principles of Ambisonics, it is a
speaker-based receiver, because encoding and decoding is combined.

All speaker based receiver types use a $\maxre$ first-order
Ambisonics decoder for decoding of diffuse sounds:
\begin{equation}
\mathbf{D} = \frac{1}{N}
\left(
\begin{array}{cccc}
  \sqrt{2} & g\tilde{s}_{1,x} & g\tilde{s}_{1,y} & g\tilde{s}_{1,z} \\
  \vdots   & \vdots           & \vdots           & \vdots           \\
  \sqrt{2} & g\tilde{s}_{n,x} & g\tilde{s}_{n,y} & g\tilde{s}_{n,z}
\end{array}\right).
\end{equation}
\(g\) is the decoder type dependent gain; for $\maxre$ this is
\(g=\frac{1}{\sqrt{2}}\) in the two-dimensional case and
\(g=\frac{1}{\sqrt{3}}\) in the three-dimensional case
\cite{Daniel2001}.

\subsubsection*{Ambisonics-based receivers}
First, second and third order receiver types were implemented. 
They follow the channel sequence and panning weight definition of the Ambisonics Association
\cite{Ambisonics2008}, using Furse-Malham normalization.
The Ambisonics-based receivers encode plane waves, i.e., they do not
account for near-field effects.
For two-dimensional encoding, all output channels which are zero, $w_n\equiv 0$, are discarded.

\subsubsection*{Binaural rendering}
Binaural signals and multi-channel signals for hearing aid microphone
arrays $\hat{\mathbf{z}}=(\hat{z}_1,\dots,\hat{z}_m)$ are generated by
rendering to a virtual loudspeaker array, i.e., using a speaker-based
render format, and applying a convolution of the loudspeaker signals
$z_n$ with the corresponding head-related impulse responses (HRIRs)
$h_{n,m}$ for the respective loudspeaker directions.
The HRIRs can be either recorded (e.g.,
\cite{Kayser2009,Thiemann2015}) or modeled
\cite{Duda1993}.

\section*{Implementation}

The implementation of TASCAR utilizes the Jack Audio Connection Kit
\cite{Davis2003}, a tool for real-time audio routing between
different pieces of software, and between software and audio hardware.
The audio content is transferred between different components of
TASCAR via JACK input and output ports.
The JACK time-line is used as a base of all time-varying features, for
data logging and as a link to the time-line of external tools.

The audio signals are processed in blocks.
Time-variant geometry and the dependent simulation coefficients, e.g.,
delay, air absorption filter coefficients or panning weights, are
updated at the block boundaries.
The simulation coefficients are linearly interpolated between the
boundaries.
This approximation by linear interpolation might be inaccurate if the
simulation coefficients vary non-linearly within a block, e.g.,
panning weights during fast lateral movements.

Render formats and algorithmic trajectory generators are implemented
as modules.
Object properties, like geometry data, reflection properties and
gains, and the time-line can be controlled interactively via a network
interface.

To achieve parallel processing in TASCAR, virtual acoustic
environments can be separated into multiple scenes.
Independent scenes can be processed in parallel.
Feedback signal paths, e.g., caused by room coupling or external
reverberation, are possible, but will lead to an additional block of
delay.
The delay and processing order of scenes is managed by the JACK audio
back-end.

\section*{Performance measurements}

For a rough estimation of the factors of computational complexity in
TASCAR, the CPU load was measured as a function of several relevant
factors.
The performance measurements were done with version 0.169 of
TASCAR. All underlying render tools are part of the TASCAR repository
\cite{tascargit}.

\subsection*{Methods}

CPU load $C$ caused by audio signal processing was assessed using the
'clock()' system function, after processing 10~seconds of white noise
in each virtual sound source.
The number of primary sources $K$, number of output channels $N$,
block size $P$, maximum length of delay lines $l_d$ and the render
format was varied (see Table \ref{tab:pspace} for an overview of the
parameter space).
No image sources were processed, i.e., all simulated sources were
primary sources, and no reflectors were used during the performance
measurements.
Each measurement of a combination of $K$, $N$, $P$, $l_d$ and render
format was repeated twice.
The CPU load $C$ is time per cycle $\tau_P$ in samples divided by
length of cycle $P$ in samples.
Number of sources and number of output channels are directly related
to the numerical complexity in the receiver module.
The block size controls the frequency of the geometry update.
Memory usage is mainly affected by the maximum delay line length.
One delay line is allocated in memory for each sound source.
At 44.1~kHz sampling rate, the memory usage of the delay lines is 520
Bytes per meter and source.
Different render formats may differ in their numerical complexity.

\begin{table}
\centering
\caption{{\bf Parameter space of the performance measurements.}}\label{tab:pspace}
\begin{tabular}{|l|l|}
\hline
Factor                          & Values                        \\
\thickhline
number of sources $K$           & 1, 10, 100, 256               \\
\hline
number of output channels $N$   & 8, 48, 128                    \\
\hline
block size $P$                  & 64, 256, 1024\,samples        \\
\hline
maximum delay line length $l_d$ & 1\,m, 10\,km                  \\
\hline
render format                   & NSP, VBAP, HOA2D              \\
\hline
CPU model                       & i5-2400@3.1GHz                \\
                                & i5-6300HQ@2.3GHz              \\
                                & i5-6500@3.2GHz                \\
                                & i7-7567U@3.5GHz               \\
%                               & i7-4930K@3.4GHz               \\
%                               & Intel Pentium E5300 @ 2.60GHz \\
                                & AMD FX-4300@3.8GHz            \\
                                & AMD Ryzen 71700               \\
\hline
\end{tabular}
\end{table}

\subsection*{Results}

A one-way analysis of variances revealed that at all tested factors
except for the delay line length and repetition showed a significant
influence on the $\tau_P$ at a significance level of $p=0.05$.
Thus, in the following analysis the data was averaged across $l_d$ and
repetitions.

\begin{figure}[!h]
%\centerline{\includegraphics[width=75mm]{figures/cpuloadexample.pdf}}
\centerline{\includegraphics[width=75mm]{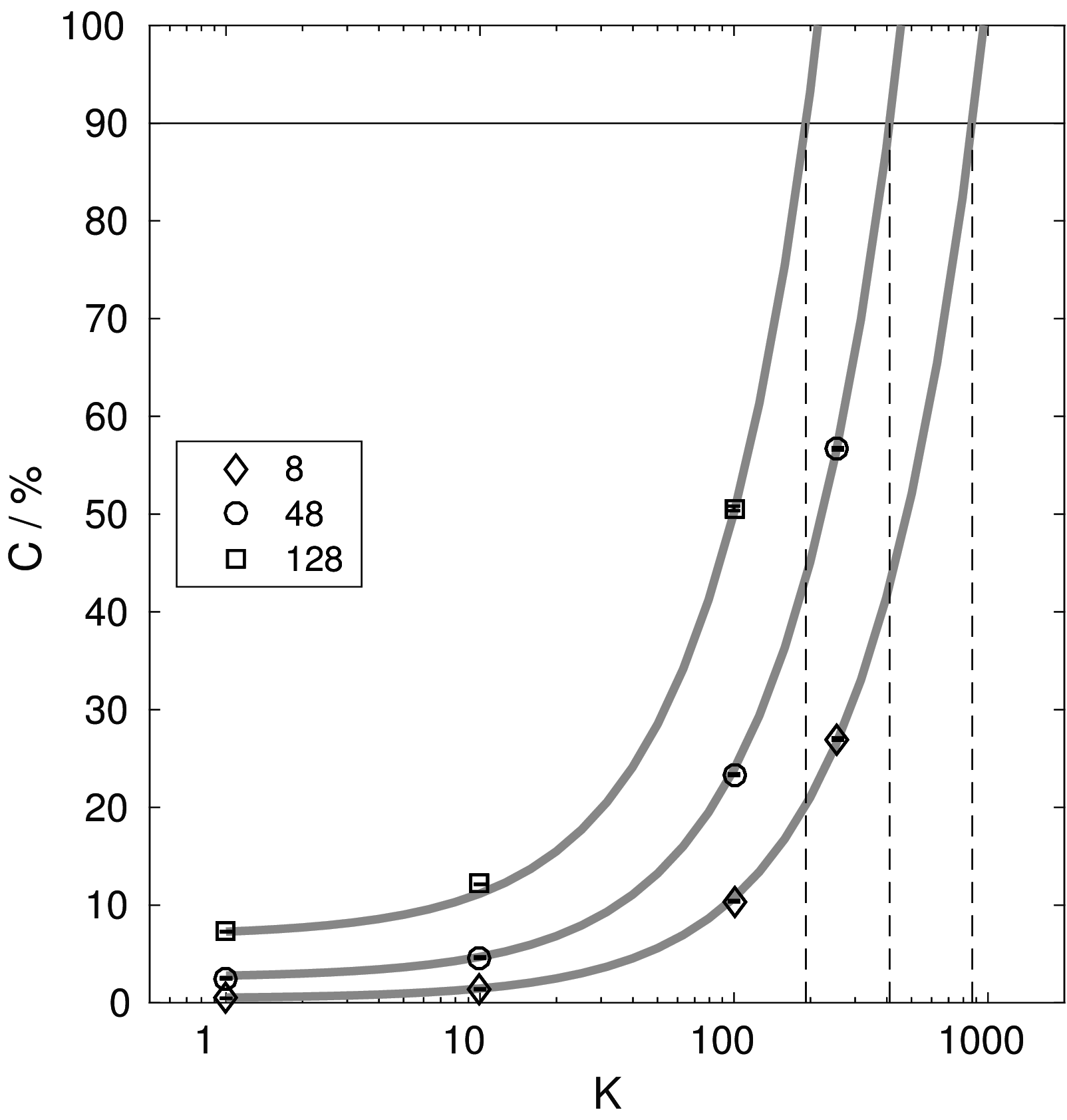}}
\caption{{\bf Example CPU load (i7-7567U@3.5GHz, HOA2D receiver,
    $P=1024$): Measured data (symbols) with model fit (Eq
    (\ref{eq:cpuload}), gray solid lines), for $N=8$\,speakers (diamonds),
    $N=48$\,speakers (circles) and $N=128$\,speakers (squares). Vertical dashed
    lines indicate the maximum possible number of sources, Eq
    (\ref{eq:kmax}), for the given hardware.}}\label{fig:cpuload}
\end{figure}

To provide an estimation of the contribution of different factors to
the numerical complexity, a model function based on the implementation
was fitted to the measured data:
\begin{equation}
\tau_P = 
\underbrace{a_0}_\textrm{overhead} + 
\underbrace{a_1 K}_\textrm{geometry} + 
\underbrace{a_2 K P}_\textrm{source audio} + 
\underbrace{a_3 N P}_\textrm{postproc.} + 
\underbrace{a_4 N K P}_\textrm{panning}.
\label{eq:cpuload}
\end{equation}
In this model, $a_0$ represents the overhead by framework which is not
related to the simulation properties.
$a_1$ is an estimate of geometry processing time, which is performed
for each source, but not depending on the number of audio samples per
processing block $P$.
The factor $a_2$ is related to source audio processing time per sample
in the transmission model, and the processing time spent in the
receiver, which does not depend on the number of speakers.
$a_3$ is an estimate of the post processing time per audio sample in
the receiver, which does not depend on the number of sources.
$a_4$ is time per audio sample for each loudspeaker and sound source,
i.e., time spent in the panning function of the render format.
The model parameters were found by minimizing the mean-square error between
the measured and predicted CPU load $C$, and are shown in Table \ref{tab:results}.
An example data set for one architecture and receiver type is shown in
Fig \ref{fig:cpuload}.

%             &        & {\scriptsize overhead} & {\scriptsize geom.} & {\scriptsize source} & {\scriptsize postpr.} & {\scriptsize panning} & {\scriptsize $P=1024$} & {\scriptsize $P=1024$} \\

\begin{table}[!ht]
%\begin{adjustwidth}{-2.25in}{0in} % Comment out/remove adjustwidth environment if table fits in text column.
\caption{{\bf Results of the model fits of CPU load measurement.}}\label{tab:results}
\begin{tabular}{|ll|ccccc|cc|}
\hline
CPU             & format & $a_0$   & $a_1$   & $a_2$   & $a_3$   & $a_4$   & $K_{\textrm{max},8}$ & $K_{\textrm{max},48}$ \\
\thickhline
%i5-2400        & NSP    & 0.81    & 1.9     & 0.13    & 0.079   & 0.0079  & 466                  & 170                   \\
%@3.1GHz        & VBAP   & 0.68    & 1.9     & 0.12    & 0.082   & 0.0077  & 486                  & 174                   \\
%               & HOA2D  & 0.8     & 1.9     & 0.14    & 0.1     & 0.0051  & 498                  & 223                   \\
%\hline
%i7-4930K       & NSP    & 0.91    & 1.9     & 0.16    & 0.16    & 0.0077  & 401                  & 155                   \\
%@3.4GHz        & VBAP   & 0.73    & 1.8     & 0.17    & 0.15    & 0.0075  & 390                  & 157                   \\
%               & HOA2D  & 0.82    & 1.9     & 0.17    & 0.18    & 0.0049  & 419                  & 200                   \\
%\hline
%AMDFX-4300     & NSP    & 0.78    & 1.9     & 0.13    & 0.17    & 0.015   & 360                  & 98                    \\
%@3.8GHz        & VBAP   & 0.79    & 1.9     & 0.15    & 0.17    & 0.014   & 340                  & 98                    \\
%               & HOA2D  & 0.95    & 1.8     & 0.23    & 0.19    & 0.0033  & 340                  & 206                   \\
%\hline
\hline
i5-2400         & NSP    & 0.045   & 0.017   & 0.001   & 0.00052 & 7.8e-05 & 541                  & 182                   \\
@3.1GHz         & VBAP   & 0.41    & 0.093   & 0.00036 & 0.00051 & 8.1e-05 & 812                  & 201                   \\
                & HOA2D  & 0.52    & 0.02    & 0.001   & 0.00088 & 4.1e-05 & 662                  & 288                   \\
\hline
i5-6300HQ       & NSP    & 0.028   & 0.0051  & 0.0011  & 0.00043 & 6.3e-05 & 548                  & 210                   \\
@2.3GHz         & VBAP   & 0.019   & 0.062   & 0.00059 & 0.00046 & 7e-05   & 742                  & 220                   \\
                & HOA2D  & 2.1e-06 & 0.016   & 0.0011  & 0.00069 & 3.7e-05 & 636                  & 302                   \\
\hline
i5-6500         & NSP    & 0.057   & 0.0034  & 0.001   & 0.00038 & 5.6e-05 & 615                  & 238                   \\
@3.2GHz         & VBAP   & 0.021   & 0.059   & 0.00053 & 0.00042 & 6.2e-05 & 825                  & 246                   \\
                & HOA2D  & 0.066   & 0.014   & 0.00098 & 0.00062 & 3.2e-05 & 714                  & 341                   \\
\hline
i7-7567U        & NSP    & 0.099   & 0.0046  & 0.00077 & 0.00036 & 4.6e-05 & 790                  & 298                   \\
@3.5GHz         & VBAP   & 0.036   & 0.071   & 0.00014 & 0.00033 & 5.2e-05 & 1443                 & 329                   \\
                & HOA2D  & 0.096   & 0.014   & 0.0008  & 0.00053 & 2.7e-05 & 868                  & 410                   \\
\hline
AMD FX-4300     & NSP    & 0.099   & 1.8e-09 & 0.00019 & 3.2e-05 & 0.00021 & 490                  & 89                    \\
@3.8GHz         & VBAP   & 1.4e-09 & 0.28    & 0.0012  & 3e-14   & 0.00017 & 316                  & 93                    \\
                & HOA2D  & 0.056   & 0.019   & 0.0016  & 0.0011  & 4.5e-05 & 441                  & 221                   \\
\hline
AMD Ryzen 71700 & NSP    & 1.1e-06 & 0.015   & 0.00087 & 0.00046 & 6e-05   & 661                  & 234                   \\
@3.6GHz         & VBAP   & 0.46    & 0.065   & 0.00027 & 0.00029 & 6.5e-05 & 1058                 & 258                   \\
                & HOA2D  & 0.064   & 0.016   & 0.00083 & 0.00061 & 3.6e-05 & 789                  & 339                   \\
\hline
\end{tabular}
%\end{adjustwidth}
\end{table}

It is often required to estimate the maximum number of sound sources
$K$ for a given CPU, render format and loudspeaker setup (affecting
$N$) and latency constraint (affecting $P$).
Eq (\ref{eq:cpuload}) can be transformed to
\begin{equation}
K_\textrm{max} \le \frac{C - a_0 P^{-1} - a_3 N}{a_1 P^{-1} + a_2 + a_4 N}.
\label{eq:kmax}
\end{equation}
As an example, $K_\textrm{max}$ was calculated for all tested
combinations of CPU model and receiver type, for $C=90$\% and
$P=1024$.
These results are given in the last two columns of Table
\ref{tab:results}, for $N=8$ and $N=48$.

The results show that on CPU models which are commonly used at the time of
writing, several hundred sound sources can be simulated.
From the tested render formats, 'HOA2D' was most efficient, especially
for larger values of $N$.
These results take only a single core into account.
On multi-core computers, more complex environments can be simulated by
splitting them into multiple environments of lower complexity, and
rendering them in parallel.

\section*{Validation and applications}

The proposed simulation tool is based on established render formats,
such as VBAP \cite{Pulkki1997} or HOA \cite{Daniel2001}.
The physical and perceptual properties of these render methods have
been extensively studied
\cite{Landone1999,Daniel2003b,Carlsson2004,Pulkki2005,Ahrens2008,Benjamin2010,Bertet2013}.
The limitations for applications in hearing aid evaluation differ from
perceptual limitations \cite{Grimm2015b}.
They depend on the sensitivity of hearing aid algorithms and the
applied hearing aid performance measures on spatial aliasing artifacts
of the render methods.
Thus the optimal render method depends on the context of a specific
application of the proposed simulation tool.
Based on the data by \cite{Grimm2015b}, a specific TASCAR scene can
be designed such that it meets the requirements of an
application-specific receiver, e.g., a human head with two-microphone
hearing aids on each ear.

Distance perception in human listeners is believed to be dominated by
the direct-to-reverberant ratio \cite{Bronkhorst1999}.
In the proposed simulation tool with a simple image source model and
position-independent externally generated late reverberation, the
distance perception may depend on simulation parameters.
Thus, in a previous study the distance perception and modeling with
room-acoustic parameters in simulations with TASCAR was evaluated
\cite{Grimm2015c}.
It was shown in a comparison of binaural recordings in a real room and
a simulation of the same geometry that in the simulation a distance
perception similar to real rooms can be achieved.

\begin{figure}[!h]
\centerline{\includegraphics[width=75mm]{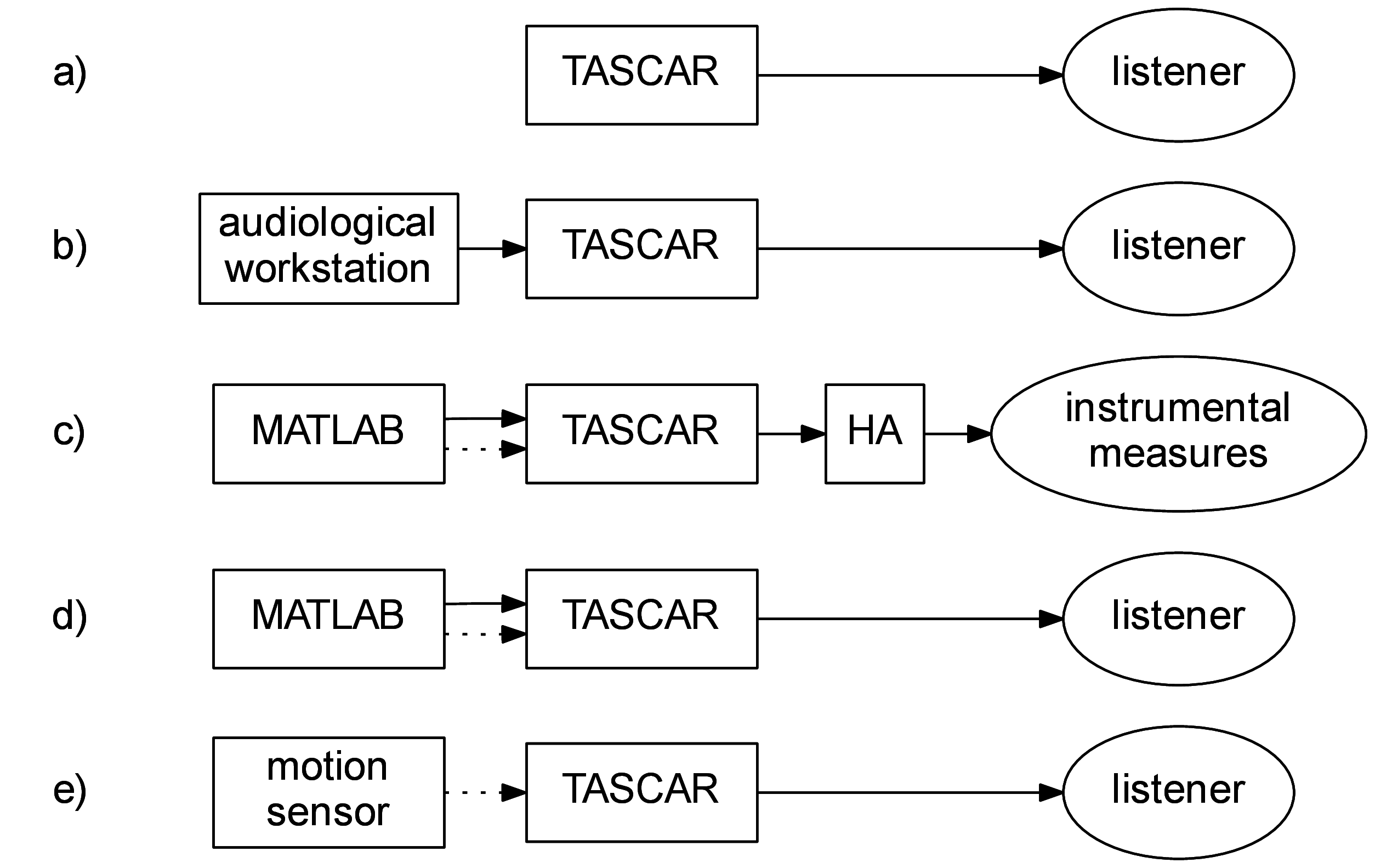}}
%\centerline{\includegraphics[width=75mm]{figures/example_applications}}
\caption{{\bf Example applications of TASCAR and its interaction. Solid
  arrows indicate audio signals, dashed arrows represent control
  information, e.g., geometry data.}}\label{fig:audapp}
\end{figure}

An overview over a number of possible applications is shown in
Fig \ref{fig:audapp}.
The simplest application of TASCAR is to play back a pre-defined
virtual acoustic environment via multiple loudspeakers (Fig
\ref{fig:audapp}.a).
For subjective audiological or psycho-acoustic measurements in virtual
acoustic environments, without hearing aids or aided with conventional
hearing aids, the audio input of virtual sound sources can be provided
by external measurement tools (Fig \ref{fig:audapp}.b).
TASCAR can also be applied to assess hearing aid (HA) performance in
simulated virtual environments, based on instrumental measures, or
with human listeners, e.g., in combination with the open Master Hearing Aid (openMHA)
\cite{Herzke2017}.
Subjective or instrumental evaluation of research hearing aids can be
performed by feeding the output of the virtual acoustic environment
directly to the inputs of a research hearing aid
\cite{Grimm2006} (Fig \ref{fig:audapp}.c).
An example study of this use case can be found in \cite{Grimm2016},
where hearing aid performance in eight different virtual acoustic
environments of different spatial complexity was assessed.
Test stimuli as well as the configuration of virtual acoustic
environment and the research hearing aid can be controlled from the
measurement platform, e.g., MATLAB or GNU/Octave
(Fig \ref{fig:audapp}.d).
Motion data can also be recorded from motion sensors or controllers,
to interact with the environment in real-time, or for data logging
(Fig \ref{fig:audapp}.e).

These use cases serve as an illustration of typical applications of
TASCAR.
The interfaces of TASCAR allow for a large number of applications.
%%
%The general structure and typical interaction with external modules is
%outlined in Fig \ref{fig:flowchart}.
%%
%
%\begin{figure}[!h]
%\centerline{\includegraphics[width=75mm]{figures/structure.pdf}}
%\caption{{\bf Schematic audio and control signal flow chart of TASCAR in a
%  typical hearing research subjective test
%  application.}}\label{fig:flowchart}
%\end{figure}

\section*{Summary and conclusions}

In this technical paper, a toolbox for creation and rendering of
dynamic virtual acoustic environments (TASCAR) was described, which
allows direct user interaction.
This tool was developed for application in hearing aid research and
audiology.
The three main modules of TASCAR - audio player, geometry processor
and acoustic model - form the simulation framework.
The audio player provides the tool with audio signals, the geometry
processor keeps track of the distribution of the objects in the
virtual space, and the acoustic model performs the room acoustic
simulation and renders the scene into a chosen output format.
The simulation uses a transmission model and a geometric image source model in
the time domain, to allow for interactivity, and for a simple physical
model of motion-related acoustic properties, such as Doppler shift and
comb filtering effects.
TASCAR allows selecting from a number of various rendering
formats, customized to the needs of a range of applications, including
higher order Ambisonics and binaural rendering formats.
%

%performance measurements
Performance measurements quantify the influence of factors related to
simulation complexity.
The results show that, despite some limitations in terms of complexity
of the virtual acoustic environment, several hundred virtual sound
sources can be interactively rendered, even over huge reproduction
systems and on consumer-grade render hardware.

It can be concluded that the proposed tool is suitable for hearing aid
evaluation.
It offers a set of features, e.g., dynamic time-domain geometric image
source model, diffuse source handling, directional sources, which is
to current knowledge unique in this combination.

\section*{Acknowledgments}
This study was funded by the German Research Council DFG FOR1732.

%\nolinenumbers

% Either type in your references using
% \begin{thebibliography}{}
% Text
% \end{thebibliography}
%
% or
%
% Compile your BiBTeX database using our plos2015.bst
% style file and paste the contents of your .bbl file
% here. See http://journals.plos.org/plosone/s/latex for 
\bibliographystyle{abbrvnat}
\bibliography{grimm}
%\include{tacar_tech_paper.bbl}

%\begin{thebibliography}{10}
%
%\bibitem{bib1}
%Conant GC, Wolfe KH.
%\newblock {{T}urning a hobby into a job: how duplicated genes find new
%  functions}.
%\newblock Nat Rev Genet. 2008 Dec;9(12):938--950.
%
%\bibitem{bib2}
%Ohno S.
%\newblock Evolution by gene duplication.
%\newblock London: George Alien \& Unwin Ltd. Berlin, Heidelberg and New York:
%  Springer-Verlag.; 1970.
%
%\bibitem{bib3}
%Magwire MM, Bayer F, Webster CL, Cao C, Jiggins FM.
%\newblock {{S}uccessive increases in the resistance of {D}rosophila to viral
%  infection through a transposon insertion followed by a {D}uplication}.
%\newblock PLoS Genet. 2011 Oct;7(10):e1002337.
%
%\end{thebibliography}

\end{document}